\documentclass[pra,twocolumn,aps,superscriptaddress,showpacs,floatfix]{revtex4}

\usepackage{amsmath}
\usepackage{amssymb}
\usepackage{amsbsy}
\usepackage{graphicx}
\usepackage{color}
\usepackage{epstopdf}
\usepackage{mathrsfs}

\begin{document}

\title{Topological phase transition and charge pumping in a one-dimensional periodically driven optical lattice}

\author{Zhihao Xu}
\affiliation{Institute of Theoretical Physics, Shanxi University, Taiyuan 030006, China}
\email{xuzhihao@sxu.edu.cn}
\author{Yunbo Zhang}
\affiliation{Institute of Theoretical Physics, Shanxi University, Taiyuan 030006, China}
\author{Shu Chen}
\affiliation{Beijing National Laboratory for Condensed Matter Physics,
	Institute of Physics, Chinese Academy of Sciences, Beijing 100190, China}
\affiliation{School of Physical Sciences, University of Chinese Academy of Sciences, Beijing, 100049, China}
\affiliation{Collaborative Innovation Center of Quantum Matter, Beijing, China}

\begin{abstract}
Experimental realizations of topological quantum systems and detections of topological invariants in ultracold atomic systems have been a greatly attractive topic. In this work, we propose a scheme to realize topologically different phases in a bichromatic optical lattice subjected to a periodically driven tilt harmonic oscillation, which can be effectively described by a superlattice model with tunable long-range hopping processes. By tuning the ratio of nearest-neighbor (NN) and next-nearest-neighbor (NNN) hopping amplitudes, the system undergoes a topological phase transition accompanied by the change of topological numbers of the lowest band from $-1$ to $2$. Using a slowly time-periodic modulation, the system emerges distinct quantized topological pumped charges (TPCs) of atoms in the filled band for different topological phases. Our scheme is realizable in current cold atomic technique.

\end{abstract}


\maketitle

\section{Introduction}
 Exploration of topological phases of matters has attracted longstanding interest in condensed matter physics in past decades \cite{Thouless,Klitzing,Kitaev,Ivanov,LiangFu,Lutchyn,Beenakker,Mourik,MTDeng,ADas,Rokhinson,Nadj-Perge,Albrecht,LJiang,Kane,Hasan,XLQi,Bernevig}. Besides traditional solid materials, ultracold atomic systems have provided a powerful platform for the investigation and simulation of topological physics \cite{Aidelsburger1,Aidelsburger2,Miyake,Jotzu,ZWu,Atala,Lohse,Nakajima,Schweizer,LeiWang,Lang}.  By manipulating the geometry of optical lattices and atomic hopping configurations, a series of celebrated models of topological quantum systems, such as the Hofstadter model \cite{Aidelsburger1,Aidelsburger2,Miyake}, the Haldane model \cite{Jotzu,ZWu} and the Su-Schrieffer-Heeger model \cite{Atala,Lohse,Nakajima,Schweizer,LeiWang}, have been experimentally realized in cold atomic systems. Besides the tunability of geometrical structures, the time-periodic modulation of parameters of systems can significantly change the band structures of the effective Floquet Hamiltonian \cite{Aidelsburger1,Aidelsburger2,Miyake,Jotzu,ZWu,JiangbinGong1,JiangbinGong2,ZhenyuZhou,DYXing,JiangbinGong3,JingZhang,Galitski,Rechtsman,Jarillo-Herrero,Kitagawa1,Rudner1} and thus provides an additional freedom to adjust nontrivially topological bands. So far, the Floquet band engineering has become an effective experimental tool for exploration of topological phases in periodically driven optical lattices.

On the other hand, the time-periodic driving has been widely applied to realize coherent manipulation of the ultracold atomic gases in the optical lattice. Some important experimental progresses include the coherent control of the single-particle tunneling amplitude in periodically shaken
lattices \cite{Lignier}, the realization of dynamical localization\cite{Lignier,Eckardt09,Creffield10}, the implementation of kinematic frustration \cite{Struck}, and the dynamical control of the quantum phase transition from a bosonic superfluid to a Mott-insulating state \cite{Zenesini}.  These experimental schemes are based on the observation of an effective modification of tunneling matrix elements induced by the time-periodic driving. In most previous studies, only the NN hopping processes are considered, due to the tunneling amplitudes between NNN sites decay very quickly and are generally negligible in comparison with the NN hopping terms. However, as we shall demonstrate in this paper, it is possible to get much stronger NNN tunneling terms for a periodically driven system. The presence of long-range hopping terms can significantly change the band structure of the system, and may induce the phase transition between topologically different phases.


In this work, we study the topological phase transition in one-dimensional (1D) periodically driven optical superlattices, which can be realized by trapping fermions in a 1D bichromatic optical lattice with a tilt harmonic oscillation. The effective time-independent Hamiltonian in the high-frequency oscillation regimes can be described by a superlattice model with both the NN and NNN hopping terms, and particularly the ratio of NN and NNN hopping amplitudes can be adjusted freely. In comparison with the 1D superlattice model described by the Harper model \cite{Lang,Kraus}, the system provides different topological behaviors when the NNN hopping amplitude $J_2$ is larger than the NN one $J_1$. To characterize the topological features, we study the topological Chern numbers of the system in different parameter regimes. The existence of the topological phase transition is based on the change of Chern numbers for the system with the lowest band being fully filled by fermions when the parameter crosses the transition point, the change of energy gaps and nontrivial edge states. Particularly, recent progress in \emph{in situ} detection with the single-site resolution offers the possibility of the detection of topological Thouless pumping \cite{Lohse,Nakajima} in different topological regimes. Our calculation verifies the topological quantization of the center of mass (CM) of the cloud in realistic ultracold atom experimental situations.

\section{Model and effective Hamiltonian}
Consider the noninteracting ultracold
fermions trapped in a bichromatic optical lattice with a tilt harmonic oscillation. The fermionic motion along the $x$ axis is described by $H=H_s+W(t),$ where
\begin{equation}\label{eq1}
H_s=-\sum_{j,m}J_m^{\prime}(\hat{c}_j^{\dagger}\hat{c}_{j+m}+h.c.)+\lambda \sum_{j} \cos{(2\pi\alpha j+\delta)}\hat{n}_j,
\end{equation}
is a bichromatic optical lattice and
\begin{equation}\label{eq2}
W(t)=2\chi \cos{(\Omega t)}\sum_j j\hat{n}_j,
\end{equation}
represents a tilt of the lattice with harmonic oscillation, where $\chi$ is the strength of the shaking term and $\Omega$ is the driving frequency. Here, $\hat{c}_{j}$ is the annihilation operator, $\hat{n}_j=\hat{c}_j^{\dagger}\hat{c}_j$ is the density operator on site $j$, $\lambda$ is the strength of the modulation, $\alpha$ determines the modulation period chosen $\alpha=1/3$ in this paper, and $\delta$ is an arbitrary phase. In the absence of the $W(t)$ term, the static bichromatic optical lattice model includes the hopping terms and the quasiperiodic modulation term where the strength of the modulation $\lambda$ depends on the amplitude of the secondary lattice and the amplitude of the hopping terms $J_m^{\prime}$ can be described by an asymptotic law $J_m^{\prime} \sim (x_m)^{-3/2}e^{-hx_m}$ \cite{Kohn,LHe}. Here, $x_m$ is the interval between two lattice sites and $h$ is the distance between the branch point and the real axis in the complex momentum $k$ space. For the deep well, $h/k_L  \sim \sqrt{V_0/(4E_r)}-1/4$, whereas in the weak binding case, $h/k_L\sim V_0/(8E_r)$ where $V_0$ is the depth of the primary lattice, and $E_r=\hbar^2 k_L^2/(2\mu)$ is the recoil energy with the wave vector of the primary laser light waves $k_L$ and $\mu$ being the mass of the fermions. With the decrease of $V_0/E_r$, the effect of the long-range hopping emerges. For the shallow potential case (take $V_0/E_r=3$ as an example), the ratio of the strengths between the NNN and the NN hopping $J_2^{\prime}/J_1^{\prime}\sim 0.1$ \cite{Zwerger,Boers}. In our work, we only consider the NN and NNN hopping terms for the shallow potential case.

\begin{figure}[tbp]
	\begin{center}
		\includegraphics[width=.5 \textwidth] {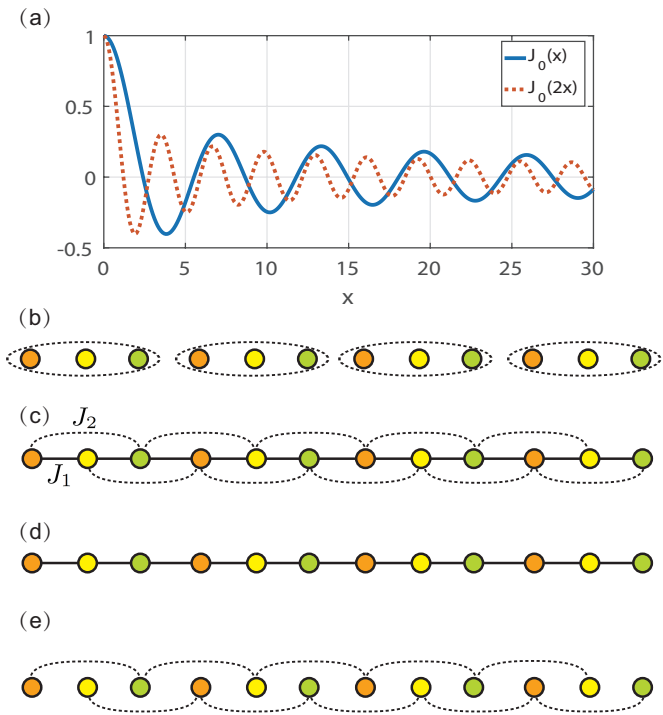}
	\end{center}
	\caption{(Color online)  (a) Bessel function of order zero $\mathrm{J}_0(x)$ and $\mathrm{J}_0(2x)$. (b) For $\alpha=1/3$, there are three different sites in every unit cell. (c) For arbitrary $\theta$, the strength of NN hopping is $J_1$ and NNN hopping is $J_2$. (d) In the $\theta=0$ limit, $J_2$ is obviated and (e) in the limit $\theta=\pi/2$, the original chain is separated into two new chains and only the NNN hopping term conserved connects odd or even sites.}\label{Fig1}
\end{figure}

In the presence of periodic shaking, the scenario has been investigated theoretically \cite{Eckardt05,Eckardt10,Hemmerich,Rahav,Goldman14,Alessio,Goldman15,Dauphin,BiaoHuang,Eckardt17} and experimentally \cite{Aidelsburger1,Aidelsburger2,Miyake,Jotzu,ZWu,JingZhang,Lignier,Eckardt09,Creffield10,Struck,Zenesini,Meinert}. For sufficiently high driving frequencies, the periodic shaking system can be equivalent to an effective Hamiltonian $H_{\mathbf{eff}}$ which behaves similarly as the undriven system, but with the hopping matrix elements $J_1^{\prime}$ and $J_2^{\prime}$ replaced by the renormalized matrix elements $J_1^{\prime}\mathrm{J}_0[2\chi/(\hbar\Omega)]$ and $J_2^{\prime}\mathrm{J}_0[4\chi/(\hbar\Omega)]$, respectively, where $\mathrm{J}_0$ is the Bessel function of order zero. Figure \ref{Fig1}(a) shows the Bessel function of order zero $\mathrm{J}_0(x)$ and $\mathrm{J}_0(2x)$. We can see the ratio of $\mathrm{J}_0(2x)/\mathrm{J}_0(x)$ can be freely controlled by $x$. In realistic experiment, we can adjust the driven strength $\chi$ and the driving frequency $\Omega$, to freely control the ratio between the strength of NNN and NN hopping terms. Hence the effective Hamiltonian can be described as the following: 
\begin{equation}\label{eq3}
H_{\mathbf{eff}}=-\sum_{j,m=\{1,2\}}J_m(\hat{c}_j^{\dagger}\hat{c}_{j+m}+h.c.)+\sum_j W_j \hat{n}_j,
\end{equation}
with
\begin{equation}\label{eq4}
W_j=\lambda \cos{(2\pi \alpha j+\delta)}.
\end{equation}
For $\alpha=1/3$, there are three different sites in each unit cell shown in Fig.\ref{Fig1}(b). In order to simplify the extra variables, we define $J_1=J\cos{\theta}$ denotes the NN hopping strength and $J_2=J\sin{\theta}$ is the NNN hopping strength [Fig.\ref{Fig1}(c)] where $J=\sqrt{\{J_1^{\prime}\mathrm{J}_0[2\chi/(\hbar\Omega)]\}^2+\{J_2^{\prime}\mathrm{J}_0[4\chi/(\hbar\Omega)]\}^2}$ and $\cos{\theta}=J_1^{\prime}\mathrm{J}_0[2\chi/(\hbar\Omega)]/J$. Without loss of generality, $\theta$ is chosen from $0$ to $\pi/2$ to change the ratio $J_2/J_1$. In the following, $J$ is set as the energy unit.

\begin{figure}[tbp]
	\begin{center}
		\includegraphics[width=0.55 \textwidth] {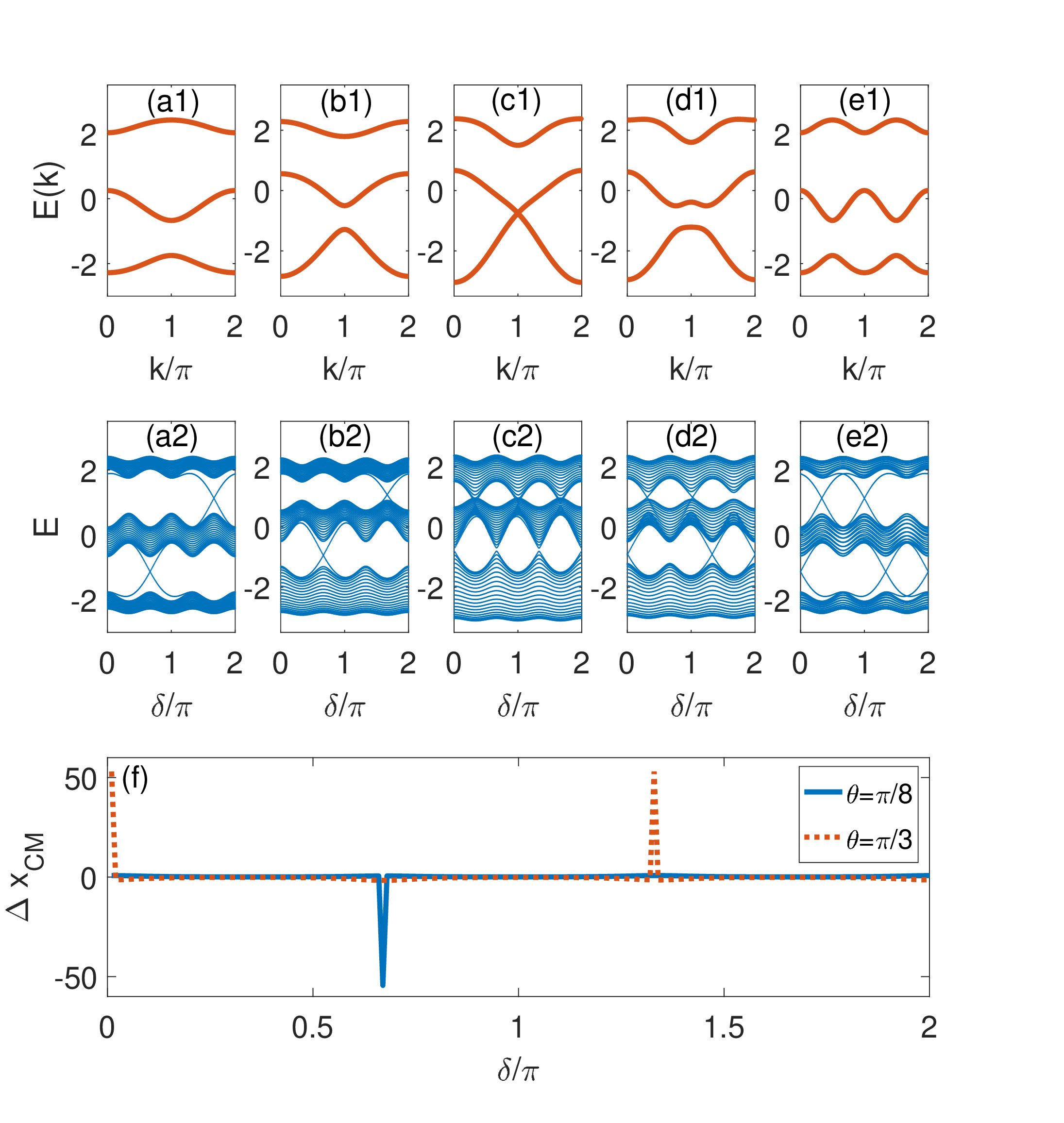}
	\end{center}
	\caption{(Color online) (a1)-(e1) Energy bands for the system under PBCs with $\lambda=1.5$, $\delta=0$, $\theta=0,\pi/8,\pi/4,\pi/3$ and $\pi/2$, respectively. (a2)-(e2) Energies varying with the phase $\delta$ under OBCs. Here, $L=60, \lambda=1.5, \theta=0, \pi/8, \pi/4, \pi/3$, and $\pi/2$, respectively. (f) The variation of the CM $\Delta x_{CM}$ vs the phase $\delta$ under OBCs with $L=60$ and filling $N=20$ fermions for the cases of $\theta=\pi/8$ and $\pi/3$.}\label{Fig2}
\end{figure}

For the case of $\theta=0$, the effective Hamiltonian Eq. (\ref{eq3}) only includes the NN hopping and the three period chemical potential terms which returns to the model studied by Lang \emph{et al.} \cite{Lang} and Kraus \emph{et al.} \cite{Kraus} [Fig.\ref{Fig1}(d)]. The system is topologically nontrivial and when the lowest band is fully filled by fermions, we can calculate the Chern number in two-dimensional phase-momentum $(\delta, k)$ space and obtain the topological number $-1$ \cite{Lang}.  Under the open boundary conditions (OBCs), the topologically nontrivial edge modes emerge in the energy gaps [Fig.\ref{Fig2}(a2)]. Whereas for the case of $\theta=\pi/2$ shown in Fig. \ref{Fig1}(e), only the NNN hopping is preserved and the $J_1$ term is omitted. The 1D chain separates into two decoupled chains, one comprises all of the odd sites and the other the even sites. These two new chains include the same hopping elements $J_2$ and the chemical potential which are still three period but with different forms from the original one, \emph{i.e.} $\lambda\cos{(\frac{2\pi}{3} j-\delta)}\hat{n}_{2j}$ and $\lambda\cos{[\frac{2\pi}{3}  (j+1)-\delta]}\hat{n}_{2j+1}$. Another difference from the original one is that the lattice constant becomes double. Hence, (i) the energy of the two new chains are degenerate; (ii) the period of the energy spectrum in momentum space shrinks half as shown in Fig.\ref{Fig2}(e1); (iii) when the lowest band is fully filled, each chain provides the Chern number $+1$ and the summation is $+2$. At the two different limits ($\theta=0$ and $\pi/2$), the system presents different topological behaviors. It means that with the increase of $\theta$ from $0$ to $\pi/2$, the system undergoes a topological phase transition with the Chern number changing from $-1$ to $+2$. In this paper, we shall study the topological phase transition with the change of $\theta$ and the charge pumping method is applied to detect the transition. We choose $\theta=\pi/8$, $\pi/4$ and $\pi/3$ as the specific examples of $\theta \neq 0$ and $\pi/2$.

\section{Topological phase transition}
 A change of topological numbers is generally attributed to a topological phase transition. As discussed above, at $\theta=0$ and $\pi/2$, the Chern number of the system are respectively equal to $-1$ and $2$. It is confirmed that with the rolling of $\theta$ from $0$ to $\pi/2$, the system undergoes a topological phase transition. To characterize the topological properties of the bulk states, we can calculate the Chern number of the lowest band in the two-dimensional parameter space of $(\delta,k)$. The Chern number is the topological invariant which is related to integral of the Berry curvature over the filled bands via $C=\frac{1}{2\pi}\int_{0}^{2\pi} d\delta \int_0^{2\pi} dk (\partial_{\delta}A_k-\partial_{k}A_{\delta})$, where $A_{X}=i\langle \psi(X)|\partial_X|\psi(X)\rangle$ is the Berry connection and $\psi(X)$ is the occupied Bloch state with the parameter $X$ \cite{Xiao}. The Chern number for fermions fully filled in the lowest band is $-1$ when $\theta$ is less than $\pi/4$, while it jumps to $2$ when $\theta$ crosses $\pi/4$. Hence the topological phase transition of the system occurs at $\theta=\pi/4$.

 If a topological phase transition exists in one system, the transition is accompanied by the energy gap closing and reopening. Next we study the change of the energy gap for different $\theta$ under the PBCs. Figures \ref{Fig2}(a1)-(e1) show the energy spectrum of Hamiltonian (\ref{eq3}) in momentum $k$ space with $\lambda=1.5$, $\delta=0$ and $\theta=0,\pi/8,\pi/4,\pi/3$ and $\pi/2$, respectively. In the absence of $J_2$ as shown in Fig.\ref{Fig2}(a1), there are three energy bands due to three different sites in each unit cell. An obvious gap emerges between the lowest two bands which is shrinking with the increase of $\theta$. As $\theta$ is reaching $\pi/4$ shown in Fig.\ref{Fig2}(c1), the first and second bands are touched together at $k=\pi$. When $\theta$ is larger than $\pi/4$, the energy gap reopens [Fig.\ref{Fig2}(d1)]. In Fig.\ref{Fig2}(e1), $\theta=\pi/2$, the 1D chain separates into two decoupled period-three new chains with the double lattice constant. The energies of two decoupled chains are degenerate and the period of the energy spectrum is shorten by half.  It is shown that $\theta=\pi/4$ is the topological phase transition point, on both sides of which one has different Chern numbers.

Under the OBCs, as the phase $\delta$ varies from $0$ to $2\pi$, the energy spectrums with $L=60$, $\lambda=1.5$, $\theta=0, \pi/8, \pi/4, \pi/3$ and $\pi/2$ are shown in Figs. \ref{Fig2}(a2)-(e2). For the case of $\theta \neq \pi/4$, the edge states connecting two different bulk regimes emerge in the gap and the position of the edge states varies continuously with the rolling of $\delta$. Whereas when $\theta=\pi/4$, the gaps vanish and there are no edge modes detected. Taking $\theta=\pi/8$ and $\pi/3$ as examples, we study the variation of the CM $\Delta x_{CM}(\delta)=x_{CM}(\delta)-x_{CM}(\delta+\delta^{\prime})$ vs the phase $\delta$ for filling $N=20$ fermions. Here, the position of CM is defined as $x_{CM}=\sum_{j}j \rho_j$, $\rho_j=\langle \psi_G |\hat{n}_j|\psi_G\rangle$ with $\psi_G$ the ground state wave function; the density distribution at site $j$ and $\delta^{\prime}$ is a tiny deviation of the phase. In Fig.\ref{Fig2}(f), for the case of $\theta=\pi/8$,  $\Delta x_{CM}\approx -54.4$ at $\delta=0.66\pi$. It means that there are fermions shifting from left edge to right with the increase of $\delta$ near $0.66\pi$. For the case of $\theta=\pi/3$, the fermions from right edge shift to left when  $\delta$ crosses $0$ and $1.33\pi$. The appearance of edge states and particles shifting from one edge to another are generally attributed to the nontrivially topological feature of bulk systems. Also the number of jump discontinuity of $x_{CM}$ is equal to the absolute value of the Chern number as show in Fig. \ref{Fig2}(f) \cite{Yan-Qing Zhu}, \emph{i.e.}, for $\theta=\pi/8$ and the lowest band fully filled, the number of the jump point is equal to the absolute value of the first band Chern number $C_1$, and for the case of $\theta=\pi/3$, the number of the jump points is $2$ corresponding to $|C_1|=2$. We can detect the jump discontinuity by an \emph{in situ} measurement of the density of the cloud, when we adiabatically pump the phase $\delta$. Thus one can directly extract Chern numbers in different regimes.

\begin{figure}[tbp]
	\begin{center}
		\includegraphics[width=0.5 \textwidth] {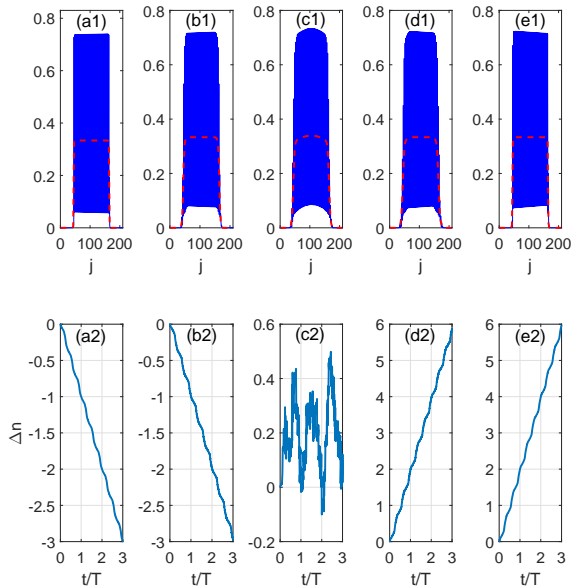}
	\end{center}
	\caption{(Color online) (a1)-(e1) The initial local density distributions of fermions $\rho_j$ labeled by blue solid lines and the initial local average densities $\bar{\rho}_j$ marked by red dash lines. (a2)-(e2) TPCs $\Delta n$ changing with time $t$. Here, we average 10 realizations of initial phase $\delta_0$, $L=210, N=40, \lambda=8.5$, and from (a)-(e) $(\theta,\omega_T)=(0,0.033), (\pi/8,0.03),(\pi/4,0.029), (\pi/3, 0.027),$ and $(\pi/2, 0.033)$, respectively. }\label{Fig3}
\end{figure}

\section{Topological pumped charges}
In realistic ultracold atom experimental situations, we consider the CM of a cloud trapped in a harmonic potential, i.e., $W_j$ in Eq. (\ref{eq4}) is replaced by
\begin{equation}\label{eq5}
W_j=\lambda\cos{[2\pi \alpha j+\delta(t)]}+\frac{1}{2}\mu \omega_T^2(j-j_0)^2,
\end{equation}
where $\delta(t)=\delta_0+2\pi t/T$ with initial phase $\delta_0$ and increasing linearly with time. The 1D lattice changes in time with a period $T$. $\omega_T$ is the frequency of the harmonic trap, $j_0$ is the position of the lattice center, and we set the mass of fermions $\mu$ as being unity. We study the density distribution of the trapped system $\rho_j$ and in order to reduce the oscillations of the density distributions, we alternatively calculate the local average density $\bar{\rho}_j=\sum_{i=1}^{p}\rho_{j+i}/p$ with $p=3$ for the case of $\alpha=1/3$.
In Figs. \ref{Fig3}(a1)-(e1), the initial local density distributions of fermions labeled by blue solid lines and the initial local average densities marked by red dashed lines are shown. Here, we average 10 realizations of the initial phase $\delta_0$, $L=210$, $N=40$, $\lambda=8.5$ and $(\theta,\omega_T)=(0,0.033), (\pi/8,0.03),(\pi/4,0.029), (\pi/3, 0.027)$, and $(\pi/2, 0.033)$, respectively. The initial local average density with a plateaus at $\bar{\rho}_j= 1/3$ in the center of the trap indicates it is a band insulator in this regime except the case of $\theta=\pi/4$. We can see the band insulators with metallic wings and the metallic edges should impact the detection of the topological numbers.


\begin{table}[htbp]
	\centering\caption{\label{I} List of the TPCs $\Delta n(T)$, the relative errors of the TPCs $\delta n(T)$ and the total particle numbers of the metallic regime $n_{\mathrm{metal}}$ for different $\theta$ and $\omega_T$ corresponding to Fig. \ref{Fig3}.  }
	\begin{tabular}{|c|c|c|c|}
		\hline \ \ \ ($\theta$, $\omega_T$) \ \ \ & \ \ \ $\Delta n(T)$ \ \ \ &\ \ \  $\delta n(T)(\%)$ \ \ \  & \ \ \ $n_{\mathrm{metal}}$ \ \ \  \tabularnewline
		\hline $(0,0.033)$     & -0.9987 & 0.13 & 0.0274 \tabularnewline
		\hline $(\pi/8,0.03)$ & -0.9921 & 0.79 & 4.0621 \tabularnewline
		\hline $(\pi/3, 0.027)$ & 1.9699 & 1.505 & 6.1032 \tabularnewline
		\hline $(\pi/2, 0.033)$ & 1.9957 & 0.21 & 1.0258 \tabularnewline
		\hline
	\end{tabular}
\end{table}

We use charge pumping method to calculate topological invariants in different parameter regimes\cite{LeeChaohong1,LeeChaohong2,LeeChaohong3}. We propose to slowly vary the phase in Eq.(\ref{eq5}) linearly with time $\delta(t)=\delta_0+2\pi t/T$, where the 1D lattices changes in time with a period $T$. To reveal the topological number, we observe the TPC $\Delta n=x_{CM}(t)/p$. Experimentally, the position of the CM can be measured by using the \textit{in situ} method. The effect of the nontrivial topological pumping can be identified as a quantization of $\Delta n$ at multiple pumping cycles. Figures \ref{Fig3}(a2)-(e2) with the same parameters as Figs. \ref{Fig3}(a1)-(e1) show the TPCs $\Delta n$ evolve with time $t$, the interval of the time is set as $\Delta t=T/10^{5}$, and average 10 realizations of the initial phase $\delta_0$. In the regime $\theta <\pi/4 $, $\Delta n(T)$ approaches to $-1$, 
whereas when $\theta$ larger than $\pi/4$, $\Delta n(T)\approx 2$. 
At the phase transition point $\theta=\pi/4$ shown in Fig. \ref{Fig3}(c2), $\Delta n(T)=0.0244$ which indicates that this point is topologically trivial.

All of the TPCs approach to integers for topologically nontrivial cases, but are not exactly equal to. We artificially define $\bar{\rho}_{j}<0.32$ as the metallic regimes. The total particle number of the metallic regime $n_{\mathrm{metal}}$ and the relative error of the TPC $\delta n(T)$ are listed in Table I. From Table I, we can see that the smaller the metallic regime, the less deviation between the measurements and the realistic topological invariants. The metallic wings will, in principle, give a non-quantized value of the pumped charge, although we could not give the definite relations between $\delta n(T)$ and $n_{\mathrm{metal}}$ (see Appendix). Hence, we shall exclude the metallic regime effect as much as possible, when we measure the topological numbers. 

\section{Summary} In summary, we demonstrate that a 1D periodically driven bichromatic optical lattice system  can be described by a superlattice model with adjustable NN and NNN  hopping terms, which exhibits the topologically nontrivial phase transition with the topological invariants from -1 to 2. Take $\theta=0$, $\pi/8$, $\pi/4$, $\pi/3$ and $\pi/2$ as specific examples to study the topological properties of the system and determine the position of the topological phase transition point. We find that when $\theta=0$ and $\pi/8$ ($\theta<\pi/4$), the topological number of the filled lowest band is -1, whereas for the case of $\theta=\pi/3$ and $\pi/2$ ($\theta>\pi/4$), the Chern number of the lowest band jumps to 2. And the topological number of $\theta=\pi/4$ is a trivial number which is topological phase transition point. We also determine the transition by calculating change of the gaps and nontrivial edge states. The charge pumping method is applied to detect the topological invariants on both sides of the transition point. The numerical results are affected by the proportion of the initial metallic wings. In the realistic ultracold atom experimental situation, we need to reduce the regime of the metallic wings as much as possible.

\section{appendix}
\subsection{Effect of harmonic potential}

In this supplemental material, we study the effect of the frequency of the harmonic potential for the detection of topological pumped charges (TPCs). For realistic ultracold atom experimental situations, we introduce a harmonic trap into the original Hamiltonian (3) in the main text given by
\begin{equation}\label{seq1}
V_j=\frac{1}{2}\omega_T^2(j-j_0)^2,
\end{equation}
where $\omega_T$ is the frequency of the harmonic trap and $j_0$ is the position of the center of the lattice. As explained by Thouless \cite{Thouless}, when the lattice is subjected to a slow and periodical time modulation, a quantization of particle transport in a one-dimensional (1D) band insulator can be detected, which is related to the topological invariant. To study the topological number, we propose to slowly change the phase linearly in time with a period $T$, $\delta(t)=\delta_0+2\pi t/T$, where $\delta_0$ is the initial phase.

\begin{figure}[tbp]
	\begin{center}
		\includegraphics[width=0.5 \textwidth] {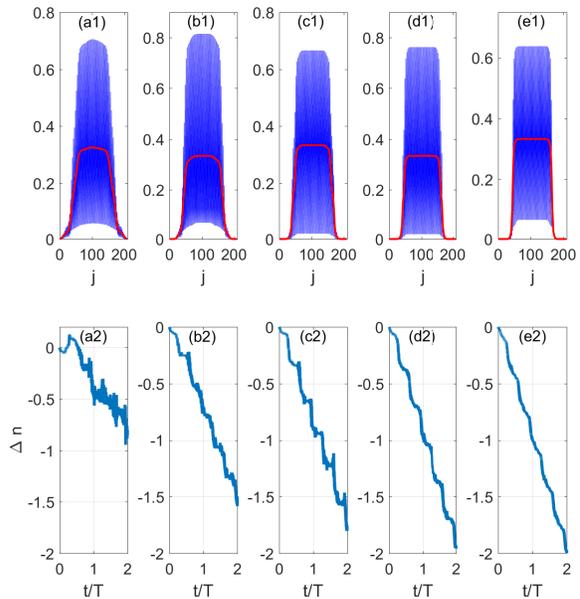}
	\end{center}
	\caption{(Color online) (a1)-(e1) The initial local density distributions of fermions $\rho_j$ labeled by blue solid lines and the initial local average densities $\bar{\rho}_j$ marked by red dashed lines. (a2)-(e2) TPCs $\Delta n$ changing with time $t$. Here, we average over 10 realizations of the initial phase $\delta_0$, $L=210, N=40, \lambda=8.5$, $\theta=0$ and from (a)-(e) $\omega_T=0.05,0.065,0.068,0.08$, and $0.1$, respectively. }\label{sfig1}
\end{figure}

In Fig. \ref{sfig1} we show the initial density distribution $\rho_j$ and the local average density $\bar{\rho}_j$ defined in the main text changing with the increasing of the frequency of the harmonic potential $\omega_T$ for the case of $L=210, N=40, \lambda=8.5$ and $\theta=0$, averaged over 10 realizations of initial phase $\delta_0$. Clearly, the local average densities present band insulators in the center of traps with metallic wings. We artificially set $\bar{\rho}_{j}<0.32$ as the metallic regime and the proportion of the total particle numbers in the metallic regime $n_{\mathrm{metal}}/N$ are listed in Table. \ref{SI}. The regime of the metallic edges shrinks rapidly with the increase of $\omega_T$. To calculate TCPs, we slowly and periodically roll the phase linearly in time and the interval of the time is set as $\Delta t=T/10^{5}$ and average over 10 realizations of initial phase $\delta_0$. Figures \ref{sfig1}(a2)-(e2) show the time evolution of TPCs $\Delta n$. For small $\omega_T$, such as $\omega_T=0.05$, $n_{\mathrm{metal}}$ is over half of total particles, and the TPC $\Delta n(T)$ at time $T$ greatly deviates the realistic topological number and the relative error of the TPC $\delta n(T)$ reaches to $55.2\%$ (shown in Table. \ref{SI}). With the increase of $\omega_T$, the metallic proportion decreases and the TPC at time $T$ approaches to $-1$ rapidly. When $\omega_T$ reaches to $0.1$, the metallic regime is much smaller, TPC $\Delta n(T)=-0.9918$,  and the relative errors of the TPC $\delta n(T)=0.82\%$. We believe to detect the topological invariants proposed by Thouless \cite{Thouless}, one need to suitably increase the tapping frequency in realistic ultracold atomic experiments to decrease the metallic regime as much as possible.

\begin{table}[htbp]
	\centering\caption{\label{SI} List of the TPCs $\Delta n(T)$, the relative errors of the TPCs $\delta n(T)$ and $n_{\mathrm{metal}}/N$ for different $\omega_T$ corresponding to Fig.\ref{sfig1}.  }
	\begin{tabular}{|c|c|c|c|}
		\hline \ \ \ $\omega_T$ \ \ \ & \ \ \ $\Delta n(T)$ \ \ \ &\ \ \  $\delta n(T)(\%)$ \ \ \  & \ \ \ $n_{\mathrm{metal}}/N$ \ \ \  \tabularnewline
		\hline $0.05$     & -0.448 & 55.2 & 0.5877 \tabularnewline
		\hline $0.065$ & -0.7231 & 27.69 & 0.3304 \tabularnewline
		\hline $0.068$ & -0.8862 & 11.38 & 0.2035 \tabularnewline
		\hline $0.08$ & -0.9713 & 2.87 & 0.1512 \tabularnewline
		\hline $0.1$ & -0.9918 & 0.82 & 0.1024 \tabularnewline
		\hline
	\end{tabular}
\end{table}

\begin{acknowledgements}
	Z. Xu thanks X. Cai for helpful discussions. This work has been supported by the NSF of China under Grants No. 11604188 and NSF for youths of Shanxi Province No. 2016021027.  Y. Zhang is supported by NSF of China under
	Grant Nos. 11234008, 11474189 and 11674201. S. Chen is supported by the National Key Research and Development Program of China (2016YFA0300600), NSFC under Grants No. 11425419, No. 11374354 and No. 11174360, and the Strategic Priority Research Program (B) of the Chinese Academy of Sciences (No. XDB07020000).
	
\end{acknowledgements}

\end{document}